\documentclass[12pt]{iopart}

\usepackage{graphicx}

\newcommand{\mycomment}[1]{\textbf{#1}}

\begin{document}

\title{Resource-efficient fibre-integrated temporal multiplexing of heralded single photons}
\author{R A Hoggarth, R J A Francis-Jones, and P J Mosley}
\address{Centre for Photonics and Photonic Materials, University of Bath, Bath, BA2 7AY, UK}
\ead{Corresponding author: r.j.a.francis-jones@bath.ac.uk}

\begin{abstract}
We present a multiplexed single photon source that re-synchronises heralded single photons generated by four-wave mixing in photonic crystal fibre using a fibre delay loop controlled by a single switch. By combining the probabilities of generating heralded single photons on four consecutive pump pulses we obtain an enhancement of the per-bin single-photon delivery probability. Our source demonstrates a way in which single-photon source multiplexing may be carried out with minimised resource overhead while retaining the benefits of a robust and alignment-free platform.
\end{abstract}



\maketitle

\section{Introduction}
Single photons are a key requirement for photonic implementations of quantum-enhanced technologies and quantum information processing. Scaling up the capabilities of algorithms and techniques currently under investigation requires the simultaneous delivery of larger numbers of single photons from independent sources \cite{Nunn2013Enhancing-Multiphoton-Rates}. These photons must be completely indistinguishable to enable high-visibility Hong-Ou-Mandel interference \cite{Mosley2008Heralded-Generation-of-Ultrafast, Lenzini2016Active-demultiplexing-of-single-photons}.

Single-photon sources are an area of vigorous research activity, and impressive progress has been made in developing sources based on single emitters such as quantum dots \cite{Somaschi2016Near-optimal-single-photon-sources, Ding2016On-Demand-Single-Photons}, colour centres \cite{Li2015Efficient-Photon-Collection}, and dye molecules \cite{Polisseni2016Stable-single-photon-emitter}. An alternative approach that minimises technical complexity is to produce photons in pairs by parametric nonlinear processes; detection of one photon can then be used to herald the presence of its twin \cite{Burnham1970Observation-of-Simultaneity-in-Parametric}. By pumping the source with short pulses and managing both the dispersion and nonlinearity of the medium, heralded single photons can be delivered with very high purity \cite{Grice2001Eliminating-frequency-and-space-time, Halder2009Nonclassical-2-photon-interference}.

Although their technical simplicity is advantageous, this type of source has the drawback that the generation mechanism is probabilistic. Hence more than one pair of photons can be generated by a single pump pulse. Despite advanced heralding detectors with photon-number-resolving capabilities, multi-pair contributions cannot be perfectly discriminated. Hence a pragmatic approach to minimising the second-order coherence of the heralded output is to reduce the generation probability until multi-pair contributions become negligible. Furthermore there exists the fundamental limit that the probability of generating exactly one pair in two single modes cannot exceed 0.25 \cite{Christ2012Limits-on-the-deterministic-creation}. While increasing pump repetition rates can help limit higher-order pair generation \cite{Broome2011Reducing-multi-photon-rates, Morris2014Photon-pair-generation-in-photonic}, ultimately this restriction on single-pair generation probability is a serious bottleneck in scaling up photonic quantum processors to operate with larger numbers of single photons simultaneously.

Conditionally routing the heralded output from several pair-generation modes into a single mode by active switching -- known as multiplexing -- provides a way of bypassing these limitations \cite{Migdall2002Tailoring-single-photon-and-multiphoton}. Multiplexing techniques can be classified by the degree of freedom used to distinguish the input modes \cite{Bonneau2015Effect-of-loss-on-multiplexed}. Spatial multiplexing schemes combine two or more spatially-distinct generation modes into an output which has a greater probability of occupation \cite{Mazzarella2013Asymmetric-architecture-for-heralded}. This approach has been investigated extensively and has been demonstrated to be effective \cite{Ma2011Experimental-generation-of-single, Meany2014Hybrid-photonic-circuit, Francis-Jones2016All-fiber-multiplexed-source, Kiyohara2016Realization-of-multiplexing-of-heralded}. However the overhead required to yield a useful increase in delivery probability is large; for example it has been shown that even in the ideal case 17 sources must be multiplexed to give over 99\% probability of delivering a single photon \cite{Christ2012Limits-on-the-deterministic-creation}. Spectral multiplexing -- enacting conditional frequency shifts to combine photons generated in different spectral modes -- has recently been demonstrated on three modes \cite{Grimau-Puigibert2017Heralded-single-photons}; larger frequency shifts will enable greater numbers of modes to be combined.

By noting that the parameter of interest is the probability of delivering a single photon per clock cycle, rather than the overall count rate, it becomes clear that temporal multiplexing, in which the outputs of a single source are re-timed and delivered at a lower repetition rate with higher occupation probability, can offer similar advantages without the need for many separate sources \cite{Schmiegelow2013Multiplexing-photons-with, Mendoza2016Active-temporal-and-spatial, Xiong2016Active-temporal-multiplexing}. The reduction in the number of usable bins per second is a price worth paying in experiments requiring many sources to fire simultaneously, as the success probability scales exponentially in probability per bin but only linearly in number of bins per second. While many schemes use concatenated networks of switches, it is significantly more resource-efficient to recycle photons through a single switch in a storage cavity \cite{Pittman2002Single-photons-on-pseudodemand, Jeffrey2004Towards-a-periodic-deterministic, Mower2011Efficient-generation-of-single, Glebov2013Deterministic-generation-of-single, Adam2014Optimization-of-periodic-single-photon, Kaneda2015Time-multiplexed-heralded-single-photon}. In this Letter we present an implementation of such a resource-efficient scheme in a fully-spliced fibre architecture using only a single source and a single multiplexing switch. The photons that we multiplex are from a source built to herald high-purity states directly, and by multiplexing over four time bins we demonstrate an improvement in the probability per bin of delivering heralded single photons.

\section{Temporal Loop Multiplexing Model}

\begin{figure}
\centering
\includegraphics[width=0.8\linewidth]{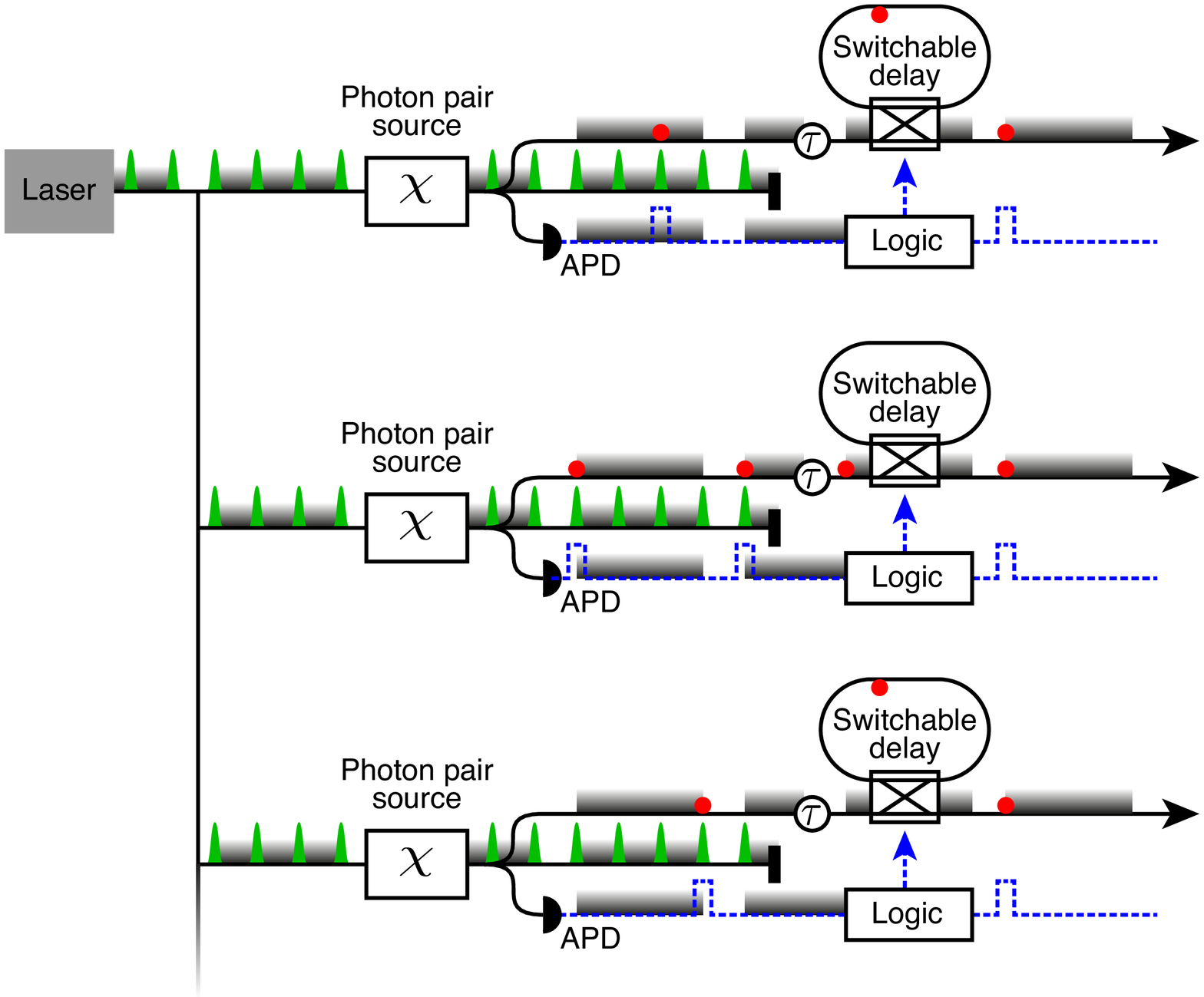}
\caption{Schematic of temporal loop multiplexing scheme. Individual sources produce heralded single photons randomly distributed between time bins. Optical delay enables feed-forward from heralding signals to control a switchable delay loop that re-times photons into every fourth time bin.}
\label{fig:schematic}
\end{figure}

A general schematic of resource-efficient temporal multiplexing using a recycling loop is shown in Figure~\ref{fig:schematic} \cite{Rohde2015Multiplexed-single-photon-state-preparation}. Our source operates at a low mean photon number per pulse, hence we can ignore contributions from multi-pair events. Considering a source that produces a thermal distribution of $n$ photon pairs per pulse with probability $p_\mathrm{th}(n)$, we define the probability that this source will successfully deliver a single heralded photon to its output as $p_1 = p_\mathrm{th}(1) \, \eta_{d}$, where $\eta_{d}$ is the lumped efficiency of the herald detection system and transmission of the remaining photon from source to output. We can the write down the probability of successfully delivering a single photon by temporal multiplexing over $m$ pulses with our loop as:
\begin{equation}
	p_1^{(m)} = 1 - \prod_{t=1}^{m}\left(1 - p_1 \, \eta_{l}^t\right),
	\label{eq:temp_multi_simple}
\end{equation}
where $\eta_{l}$ is the lumped efficiency for a single pass of the switch and storage loop. Eq.~\ref{eq:temp_multi_simple} can be thought of as the probability that all of the previous pulses do not fail to deliver a photon.

In order to compare the multiplexed source with a single source, we define an improvement factor,
\begin{equation}
f_{p1}^{(m)} = \frac{p_1^{(m)}}{p_1},
\end{equation}
which describes the change in probability of delivering heralded photons per output time bin effected by multiplexing over $m$ pulses relative to using only a single pump pulse. Of course, this does not take into account the reduction in the number of time bins resulting from multiplexing in the time domain. For a laser with repetition rate $R$ the number of output bins per second is $R/m$ and the resulting overall single-photon count rate per second is
\begin{equation}
C_1^{(m)} = \frac{R \, p_1^{(m)}}{m}
\end{equation}
yielding a change in count rate relative to a non-multiplexed (simplex) source, $C_1 = R \, p_1$ of
\begin{equation}
f_{C1}^{(m)} = \frac{C_1^{(m)}}{C_1} = \frac{p_1^{(m)}}{m \, p_1}.
\end{equation}
Finally, we consider the requirement for $N$ sources to fire simultaneously, as required in photonic quantum processors. If each source is multiplexed over $m$ pulses, the probability of delivering $N$ single photons from independent sources in the same output time bin is $p_N^{(m)} = [p_1^{(m)}]^N$ at a rate of \begin{equation}
C_N^{(m)} = \frac{R}{m} \, p_N^{(m)}.
\end{equation}
Hence we find the improvement factor expected from $m$-pulse multiplexed sources relative to non-multiplexed sources is
\begin{equation}
f_{N}^{(m)} = \frac{C_N^{(m)}}{C_1^{(m)}} = \frac{1}{m}\left(\frac{p_1^{(m)}}{p_1}\right)^N,
\end{equation}
so we see that the ``speed-up'' expected as a result of multiplexing increases exponentially in the number of independent sources $N$ as long as $p_1^{(m)} > p_1$.

\begin{figure}
\centering
\includegraphics[width=\linewidth]{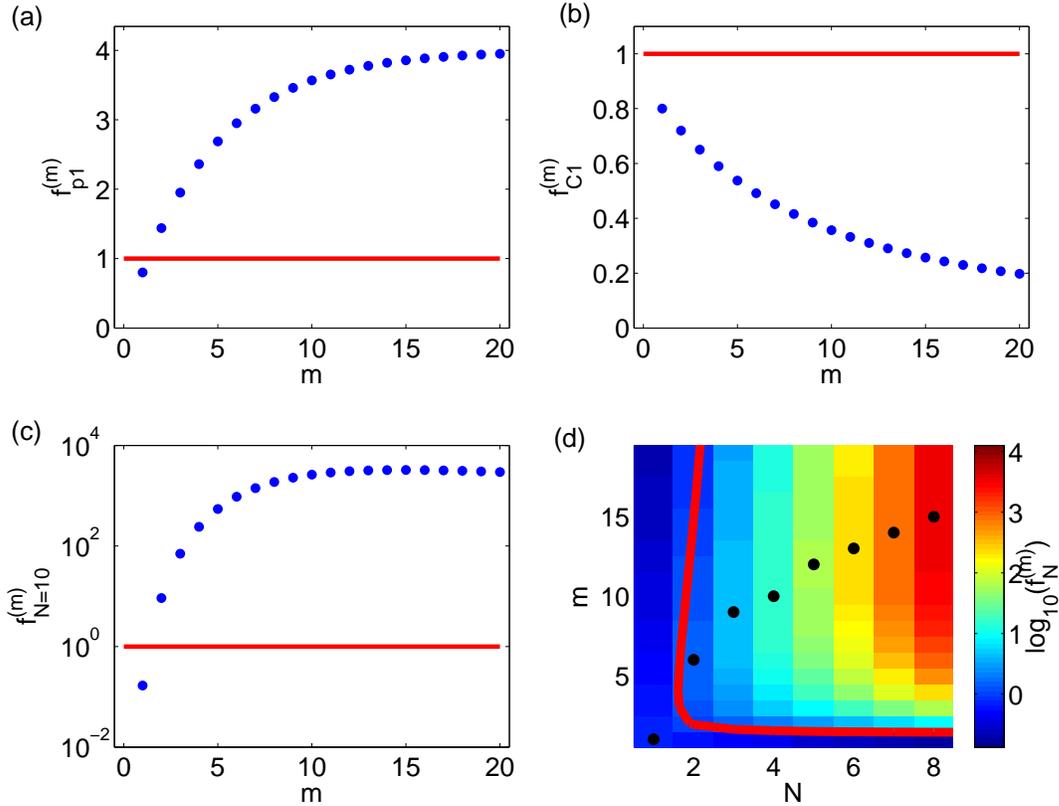}
\caption{Performance of loop multiplexing for switch loss of 1.0\,dB ($\eta_l = 0.8$) and multiplexing over up to $m = 20$ pulses. Red lines indicate break-even threshold relative to simplex source performance. (a) Enhancement in per-bin probability, $f_{p1}^{(m)}$; (b) Reduction in per-second count rate, $f_{C1}^{(m)}$; (c) Enhancement in N-source rate, $f_{N}^{(m)}$, with \mycomment{$N = 10$}; (d) Enhancement in N-source rate, $f_{N}^{(m)}$, as a function of number of sources $N$ where black circles indicate optimum value of $m$ for each $N$. Note logarithmic scaling of panels (c) and (d).}
\label{fig:theory}
\end{figure}

These quantities are plotted in Figure \ref{fig:theory} for values of repetition rate and loss that reflect our experiment. Figure \ref{fig:theory}\,(a) shows up to a factor of four enhancement in per-bin probability, $f_{p1}^{(m)}$, by multiplexing over 20 pulses, but that even with $m = 4$ the enhancement is already greater than a factor of two for our switch loss of 1\,dB. Despite the reduction in overall count rate, Figure \ref{fig:theory}\,(c) demonstrates that the scheme yields orders of magnitude improvement when considering the requirement for \mycomment{$N = 10$} sources to fire simultaneously. Finally, Figure \ref{fig:theory}\,(d) shows that loop multiplexing is advantageous in almost all multi-source experiments.

\section{Implementation}

Our starting point was a source that generates pairs of photons by four-wave mixing in photonic crystal fibre (PCF) \cite{Francis-Jones2016All-fiber-multiplexed-source}. The source was pumped by $\sim$1\,ps duration pulses at 1064\,nm derived from a 10\,MHz amplified modelocked fibre laser (Fianium FP-1060-PP) pulse-picked to 5\,MHz. The dispersion of the PCF was designed to produce heralded single photons at 1550\,nm directly in pure quantum states by minimizing frequency correlation by group-velocity matching \cite{Garay-Palmett2007Photon-pair-state-preparation}. The PCF was spliced into a fully-integrated source that separated 810\,nm signal and 1550\,nm idler photons and isolated them from the pump with fibre Bragg gratings and bandgap-guiding fibres. The 810\,nm photons were sent to a silicon avalanche photodiode to provide a heralding signal.

Our source included a 2$\times$1 optical switch to minimise noise in the output mode \cite{Brida2011Experimental-realization-of-a-low-noise}; this did not play any role in multiplexing. Using feed-forward from field-programmable gate array (FPGA) logic, the state of the switch was conditioned to be open only when the heralding detector had fired. Due to the time required for the logical operations and for the optical switch to settle into a new state, 200\,ns of static fibre delay was included between the source and the switch.

\subsection{Delay loop and loss budget}

\begin{figure}
\centering
\includegraphics[width=0.8\linewidth]{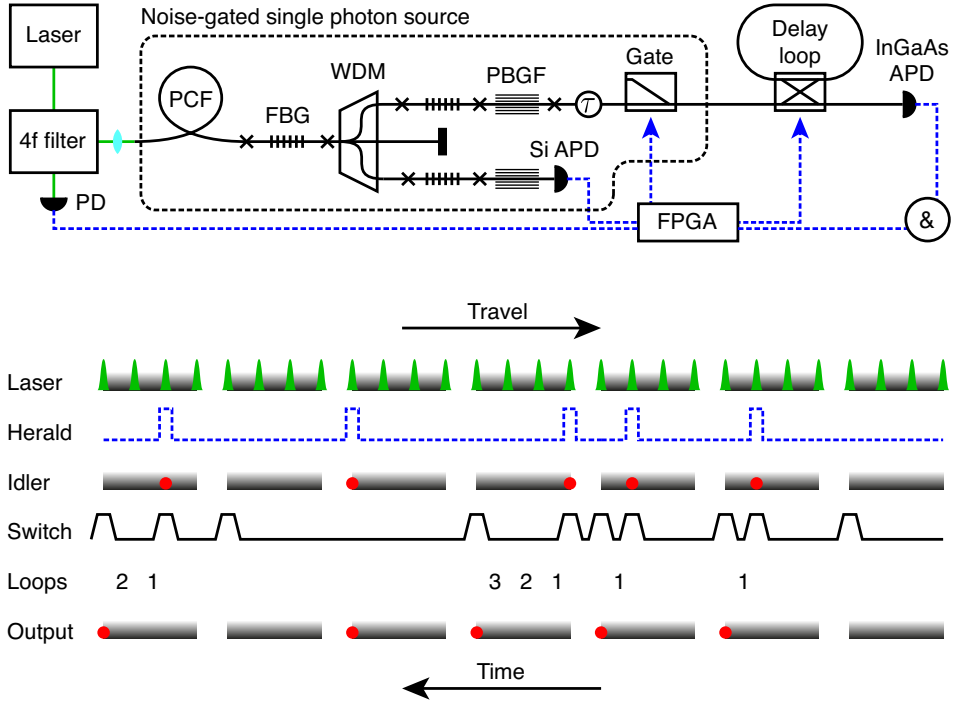}
\caption{Schematic of the source and temporal multiplexing implementation. PD - photodiode to provide clock signal; PCF - dispersion-engineered photonic crystal fibre for FWM pair generation; FBG - fibre Bragg gratings to reject pump; WDM - wavelength division multiplexer splits 1550\,nm photons (top rail) and 800\,nm photons (bottom rail) from residual 1064\,nm (centre); PBGF - photonic bandgap fibre filters for additional wavelength isolation around FWM; Si APD - silicon avalanch photodiode makes heralding detections; $\tau$ - fixed optical delay in SMF-28; FPGA - field-programmable gate array implements fast logic operations; Gate - $2 \times 1$ fibre-coupled optical switch to prevent uncorrelated noise in the 1550\,nm channel from exiting the source; Delay loop - fibre delay matched to laser repetition rate with $2 \times 2$ optical switch for multiplexing; InGaAs APD to monitor output; \& - coincidence counting.}
\label{fig:optical_scheme}
\end{figure}

To follow the source, we built a fully fibre-integrated temporal multiplexing system from a switchable loop of fibre that produces a delay similar to the pulse separation of the pump laser system (200\,ns). The delay loop, shown in Figure \ref{fig:optical_scheme}, could be inserted or removed from the optical path by a fibre-integrated 2$\times$2 optical switch. In the ``cross'' state, photons leaving the source were stored in the loop, and any light already in the loop was directed to the output. With the switch set to ``bar'', light exiting the source bypassed the loop, while light in the loop made additional passes unless removed by loss. In this manner photons could be delayed by multiples of the loop transit time while experiencing commensurate delay-dependent loss.

Due to the concatenated effect of consecutive passes through the loop, the performance of this scheme is critically dependent on switch and fibre loss. Hence it was vital that our source produce heralded photons at 1550\,nm enabling us to implement our delays in telecoms fibre (Corning SMF-28) with negligible loss. The quoted insertion loss for the switch was 1\,dB, though we measured small variations depending on the path taken. A photon heralded in a given time bin experiences loss on each pass through the loop; hence earlier time bins contribute less to the overall output relative to later bins, and total loop loss determines the number of time bins over which it is worthwhile multiplexing. In our case, the loop was used to multiplex over four consecutive pump pulses and hence deliver heralded photons in every fourth time bin.

\subsection{Feed forward control}

The control protocol is represented in Figure \ref{fig:optical_scheme}. The 5\,MHz pulse train of the pump laser was monitored with a photo-diode and used as the clock input to a phase-locked loop providing the fundamental clock for the FPGA electronics. A counter was set to label clock edges from 1 (early) to 4 (late) cyclically, and additional higher-frequency clocks were derived for the purposes of edge sampling.

When a herald signal arrived in a particular time bin, the switch was set to ``cross'' to divert the corresponding idler photon into the loop. After half the loop delay, the switch was set to ``bar'' to keep the idler circulating and a flag set to record that the loop contained a photon. At the onset of every fourth time bin, the switch was returned to ``cross'' to deliver the contents of the loop if a photon was flagged to be present.

Although not shown schematically, in the event that additional herald signals occur during one cycle of four time bins, the logic is configured to use a later-arriving photons that will be subject to lower loss. To make way for the newer photons, older photons stored in the loop are dumped to the output and exit at an incorrect time (not on the subsequent fourth time bin). This behavior adds an accidental output that can be straightforwardly gated out later; however, the frequency of such events was negligible in the experiments presented here.

Alignment of the timing signals, critical for the success of this scheme, was performed externally to the main FPGA logic by home-built electrical delay lines with up to 130ns delay available in steps of 250ps.

\section{Multiplexing performance}

Firstly we quantified the contribution of each of the four time bins to the overall count rate of heralded single photons by using an oscilloscope to create a histogram of InGaAs detector click arrival times. The results are displayed in in Table \ref{tab:contributions}. It can be seen clearly that later time bins make a much greater contribution to the overall output, as the effectiveness of earlier bins is reduced by loss. We see that the earliest bin contributes less than 10\,\% of the photons delivered; hence in our case it is clear that multiplexing over a larger number of bins would not be advantageous.

\begin{table}
\centering
\caption{Contributions to output from each time bin. The photons generated were less than 1\,ns in duration and not close to the time bin edges. The slight reduction in the time window for bin 4 is an artefact of the reset logic and is inconsequential to the operation of the source.}
\begin{tabular}{ccc}
\hline
Herald time bin & Bin width & Relative contribution \\
\hline
$b = 4$ & 121.5ns & 0.37 \\
$b = 3$ & 121.5ns & 0.35 \\
$b = 2$ & 121.5ns & 0.19 \\
$b = 1$ & 89.0ns & 0.09 \\
\hline
\end{tabular}
  \label{tab:contributions}
\end{table}

Secondly we assessed the performance of our multiplexing scheme through measuring the count rates in the target bin with and without loop multiplexing enabled. This was achieved by deactivating the switch (leaving it in the ``bar'' state) rather than making any change to the logic to avoid any electronic timing change between enabling and disabling multiplexing. Hence, the constant factor of static switch loss remains in the simplex source data but can be backed out later.

\begin{figure}
\centering
\includegraphics[width={\linewidth}]{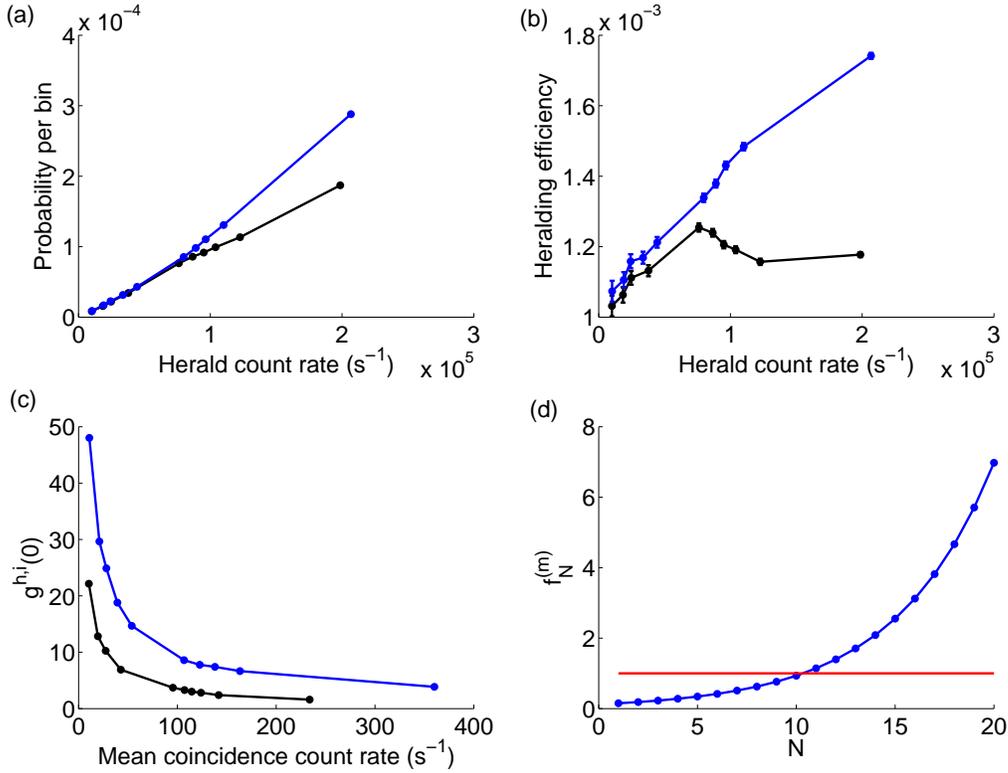}
\caption{Data displayed with loop disabled (black) and with loop multiplexing enabled (blue). (a) Probability per time bin of detecting a single photon in the output. (b) Heralding efficiency both as a function of heralding count rate. (c) Cross-correlation between herald and idler output channels. (d) Improvement in $N$-source clock rate afforded by temporal multiplexing, $f^{(m)}_N$, when per-bin probabilities are at maximum values available in panel (a). Break-even point when multiplexed sources overtake simplex sources shown in red.}
\label{fig:loopdata}
\end{figure}

Shown in Figure \ref{fig:loopdata}\,(a) is the critical metric: the probability of detecting a photon per output time bin. It can be seen at higher count rates that the loop increases the probability of detecting a photon in the target bin, indicating that we have successfully stored heralded photons and retrieved them at the correct time. The maximum improvement in probability that we measured as a result of re-timing photons with the loop was a factor of 1.54(1). In Figure \ref{fig:loopdata}\,(b), we see that loop multiplexing enables the heralding efficiency (Klyshko efficiency) to increase as the heralding count rate increases as more photons are stored and re-synchronised by the multiplexing scheme. This is in contrast with the simplex source whose heralding efficiency cannot increase beyond the limits set by the single source performance and channel loss. Figure \ref{fig:loopdata}\,(c) shows the increase in cross-correlation, $g^{\mathrm{h,i}}(0)$, between the herald and idler channels as a result of re-timing photons with the loop, defined as:
\begin{equation}
g^{\mathrm{h,i}}(0) = \frac{R C}{S_{\mathrm{h}} S_{\mathrm{i}}},
\end{equation}
where $C$ is the coincidence count rate, and $S_{\mathrm{h}}$ and $S_{\mathrm{i}}$ are the singles count rates for the re-timed herald and idler respectively. All of these figures of merit improve as a result of implementing our time-multiplexing scheme.

Although the improvement in delivery probability achieved by our system are relatively modest, considered in the context of operating several single-photon sources in parallel the benefit becomes apparent. Figure \ref{fig:loopdata}\,(d) shows the rate at which $N$ independent multiplexed sources would fire simultaneously relative to $N$ simplex sources, $F_N^{(m)}$ based on our measured source performance for a multiplexing depth of $m = 4$. In order to make a fair comparison, we have backed the static switch loss of approximately 1\,dB out of the simplex source data to yield an improvement factor of 1.22(1) at the maximum performance of our system. We see that, even for this small improvement, the break-even point at which multiplexing becomes advantageous occurs at $N = 10$ sources. With lower switch loss, the improvement would increase dramatically.

\section{Conclusion}

Through combining a switchable fibre delay loop with a PCF-based source of high-purity heralded single photons, we have demonstrated that the probability of delivering a single photon per bin can be enhanced in an integrated architecture by combining four time bins with only a single optical switch. The reduction in the number of usable bins per second is a price worth paying in experiments requiring many sources to fire simultaneously, as the success probability scales exponentially in probability per bin but only linearly in clock speed. Although the improvement we have demonstrated is modest, advances in fibre-integrated optical switch technology -- in particular reduction of loss -- will enable much larger numbers of bins to be combined while retaining the benefits of an alignment-free package.

\section{Acknowledgements}

This work was funded by the UK EPSRC Quantum Technology Hub \textit{Networked Quantum Information Technologies}, grant number EP/M013243/1. RAH thanks Hiroko and Jim Sherwin for funding.



\end{document}